\documentclass[10pt,a4paper]{article} 
\usepackage[dvips]{graphicx}
\usepackage{amsfonts,amsmath,amssymb}
\usepackage{hyperref}

\begin{document}
\title{An Investigation of the Casimir Energy for a Fermion Coupled to the Sine-Gordon Soliton with Parity Decomposition}

\author{S.S. Gousheh\footnote{Electronic address: ss-gousheh@sbu.ac.ir}, A. Mohammadi\footnote{Electronic address: a$\_$mohammadi@sbu.ac.ir}~ and L. Shahkarami\footnote{Electronic address: l$\_$shahkarami@sbu.ac.ir}\\
 \small   Department of Physics, Shahid Beheshti University G.C., Evin, Tehran
19839, Iran}
\maketitle
\begin{abstract}
We consider a fermion chirally coupled to a prescribed pseudoscalar field in the form of the soliton of the sine-Gordon model 
and calculate and investigate the Casimir energy and all of the relevant quantities for each parity channel, separately. 
We present and use a simple prescription to construct the simultaneous eigenstates of the Hamiltonian and parity in the continua from the scattering states. 
We also use a prescription we had introduced earlier to calculate unique expressions for  the phase shifts and check their consistency with both the weak and strong forms of the Levinson theorem.
In the graphs of the total and parity decomposed Casimir energies as a function of the parameters of the pseudoscalar field distinctive deformations appear whenever a fermionic bound state energy level with definite parity crosses the line of  zero energy.
However, the latter graphs reveal some properties of the system which cannot be seen from the graph of the total Casimir energy.
Finally we consider a system consisting of a valence
fermion in the ground state and find that the most energetically favorable configuration is the one with a soliton of winding number one, and this conclusion does not hold for each parity, separately.
\end{abstract}

\section{Introduction}
In 1948 Casimir in two seminal papers \cite{casimir1,casimir2} calculated the force between two uncharged perfectly conducting plates and
found that there exists a net attractive force between the metallic plates.
The Casimir effect arises from the change in the zero-point energy of the system \cite{vaccas1,vaccas2,vaccas3}.
This change could happen when nontrivial boundary conditions are imposed or nontrivial spatial background fields such as solitons are present. We shall henceforth refer to these cases as nontrivial cases.
\par
The Casimir energy and the resulting forces have been investigated for different fields in different geometries and boundary conditions \cite{geometries}.
In some of these investigations the Casimir forces on the boundaries are also calculated.
The results show that the sign of the force depends on the type of the field considered, the geometry of the problem,
the boundary conditions imposed, and the number of space-time dimensions considered (see \cite{valuyan} and the references therein).
The fact that the vacuum energy of quantum fields diverges both in the free space and nontrivial cases
have been known for many years \cite{inftyofvac}.
Many different regularization schemes have been used to properly subtract the energy of the free space vacuum
from the energy of the vacuum in the nontrivial case to find the Casimir energy.
Sometimes instead of these direct subtraction methods, various analytic continuation schemes are used,
the most popular of which is the zeta function method \cite{techniques}.
\par
Quantitative experimental confirmation of the Casimir effect became possible after the advent of the high precision probes in the 1990s.
In 1997 Lamoreaux \cite{Lamorea1,Lamorea2} conducted the first successful measurement for the Casimir effect using a
plate and a part of a conducting sphere with a large radius.
Since then, many experimental investigations have measured the Casimir force for systems with different shapes and boundary conditions \cite{exper}. 
Recently, it has been shown that the Casimir force can be utilized in the design of noncontact wear-proof parts of MEMS and NEMS \cite{mems1,mems2,mems3,mems4,mems5}.
\par
As mentioned above, the zero-point energy can also be affected by the presence of nontrivial background fields which are usually chosen to be soliton.
Also sometimes a very simple potential such as an electric potential well is chosen as the background field.
This simple choice renders the problem of vacuum polarization and the Casimir energy exactly solvable and reveals finer details of these effects \cite{dehghan}.
Some authors compute the Casimir energy as the lowest order quantum correction to the mass of the solitons
including supersymmetric solitons \cite{dashen,cassol,super1,heavy3,heavy4,la}.
Most of the models with solitons are not exactly solvable and the Casimir energy cannot be calculated directly.
In such cases, one has to resort to numerical methods which sometimes can be facilitated by the use of the indirect methods
such as the phase shift method which relates the derivative of the phase shift with respect to the momentum
to the spectral deficiency in the continuum states \cite{dehghan,dashen,super1,heavy3,heavy4,la2}.
\par
In this paper we choose a Lagrangian describing a Fermi field coupled to a prescribed pseudoscalar background field, in ($1+1$) dimensions.
The background field is chosen to be the soliton of the sine-Gordon model.
It is well known that the back-reaction of the fermion on the soliton is small and it changes the shape of the soliton very slightly \cite{jackiw,leila}.
This has motivated us to investigate this coupled fermion-soliton model with the sine-Gordon soliton as a prescribed background field.
Since the equations of motion of the fermion for our model are not analytically solvable,
we solve them using numerical methods and obtain both bound and continuum scattering states for the Fermi field.
 In our model, parity is an exact symmetry of the system.
This gives us the opportunity to investigate and explore the properties of the Casimir energy in each parity channel, separately.
The bound states are automatically parity eigenstates, since they turn out to be nondegenerate.
For the continuum states, we first compute the scattering states and then use a very simple and straightforward prescription to compute the parity eigenstates. 
For an alternative method to construct the parity eigenstates from the scattering states see \cite{heavy3,heavy4}.
In this paper, we compare our results with those of a similar model in which the prescribed pseudoscalar field
has a simple piecewise linear form which behaves like a soliton.
This simple form makes the problem exactly solvable \cite{dr}.
We have computed the Casimir energy of this Simple Exactly Solvable Model (SESM) by subtracting directly the vacuum energies
of the system in the presence and absence of the disturbance \cite{la}. 
Moreover, we have already solved an analogous problem with the kink of the $\lambda \Phi^4$ as
the topologically nontrivial prescribed background field, but without exploring the properties of the system for each parity, separately \cite{la2}. We have adjusted the parameters
of the three models to be the same so that
the results of the three models are comparable, and we compare them whenever possible.
The comparison between these models helps us to investigate how the functional form of the background field and the details of
its shape affect the results for each parity channel, as well as the scattering problem. We have also utilized SESM as a testing ground for our numerical methods.
In this paper we use the phase shifts to compute the Casimir energy. 
The phase shifts of the upper and lower components of Dirac spinors coupled to topologically nontrivial configurations are not generically equal. 
We have devised a prescription to define a unique phase shift for the Dirac spinors in these cases: 
We define the phase shift to be the average of those two phase shifts \cite{levinsondr}. 
We have checked \cite{la2,levinsondr,farid} the correctness of this prescription in several different cases using both the weak \cite{levinson} and strong \cite{levinsondr} forms of the Levinson theorem. 
In this paper we also check the consistency of our results with the Levinson theorem for the phase shifts of the parity eigenstates and also the scattering states.
\par
In section II we introduce the model.
In section III we find the bound and continuum states and plot samples of them.
In section IV we focus on the calculation of the Casimir energy by the use of phase shift method.
Also, we check the consistency of the resulting phase shifts with the weak and strong forms of the Levinson theorem.
We then plot the total Casimir energy as well as the ones in each parity channel separately, all as a function of the parameters of the pseudoscalar field.
In section V we investigate the stability of a system which consists of a valence fermion present in the ground state,
in the presence of the sine-Gordon soliton. Finally, in section VI we conclude with a brief summary.
\section{A fermion in the presence of the prescribed sine-Gordon soliton}
The model that we study in this paper consists of a Fermi field coupled to a prescribed pseudoscalar field, in the form of the soliton of the sine-Gordon model, as defined by the following Lagrangian
\begin{equation}\label{e1}\vspace{.2cm}
 {\cal L}=\bar{\psi}\left(i\gamma^{\mu}\partial_{\mu}-M \mathrm{e}^{i \phi(x)\gamma^5}\right)\psi,
\end{equation}
where the external field $\phi(x)$ is chosen to be
$\phi(x)=(m/\sqrt{\lambda})\left\{\tan^{-1} \left[\textrm{exp}(m x)\right]-\pi\right\}$.
The parameters $M$ and $m$ refer to the masses of the Fermi and pseudoscalar fields, respectively.
Our purpose is to compute the Casimir energy of this system.
The Casimir energy in the standard approach is obtained by computing the change in the zero-point energy of the field
due to the presence of the boundary conditions or the background field.
This energy for a Fermi field can be written as follows (see for example \cite{la,la2})
\begin{align}\label{e2}
E_{\mathrm{Casimir}}=&\int_{-\infty}^{+\infty}\text{d}x\int_{0}^{+\infty}\frac{\text{d}p}{2\pi}\sum\limits_{j=\pm}\left(-\sqrt{p^2+M^2}\right)\nu_p^{j\dag}\nu_p^j+
\int_{-\infty}^{+\infty}\text{d}x\sum\limits_{i}\left(E_{\text{bound}}^{i-}\right)
\chi_{2\text{b}_i}^\dag\chi_{2\text{b}_i}\nonumber\\
-&\int_{-\infty}^{+\infty}\text{d}x\int_{-\infty}^{+\infty}\frac{\text{d}k}{2\pi}\left(-\sqrt{k^2+M^2}\right)v_k^\dag
v_k\nonumber\\=&\sum\limits_{i}E_{\text{bound}}^{i-}
-\int_{0}^{+\infty}\text{d}k\sqrt{k^2+M^2}\left[\rho^{\mathrm{sea}}(k)-\rho_0^{\mathrm{sea}}(k)\right]+\frac{M}{2}.
\end{align}
The functions $\nu_p^j(x)$ and $v_k(x)$ refer to the normalized fermion wave functions for the continuum states with negative energy in the
presence and absence of the disturbance, respectively. 
The functions $\chi_{2\text{b}_i}(x)$ are the normalized fermion wave functions for
the bound states with negative energy and $E_{\text{bound}}^{i-}$ denote their energies.
The $M/2$ term in the last line takes into account the contribution of the fermionic negative parity half-bound state at $E=-M$ in the free case.
The factor $[\rho^{\mathrm{sea}}(k)-\rho_0^{\mathrm{sea}}(k)]$ is the difference between the density of the continuum states with the negative
energy in the presence and absence of the disturbance. 
One usually refers to the calculation of the Casimir energy using Eq.\,(\ref{e2}) as the direct approach.
\par
\begin{center}
\begin{figure}[h] \hspace{4.cm}\includegraphics[width=8.cm]{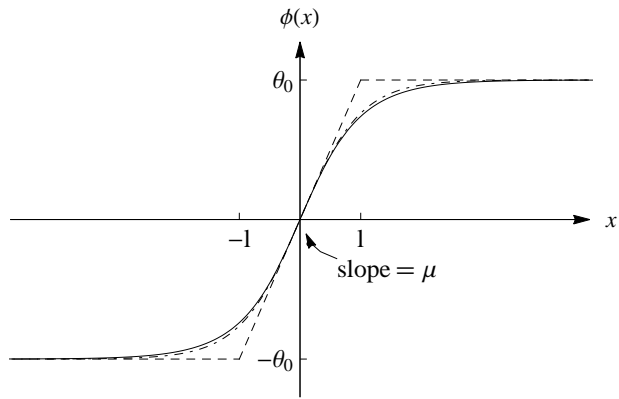}\caption{\label{sgtoy} \small
Solid, dashed and dotdashed lines show $\phi(x)$ for the sine-Gordon soliton, the simple exactly solvable model and the kink, respectively. The parameters $\theta_0$, $\mu$ and $l$ are also shown in the figure.}
\end{figure}
\end{center}

For exactly solvable systems one can calculate all the normalized continuum wave functions with negative energy in the presence of the disturbance,
$\nu_p^j(x)$, and all the negative energy bound states, $\chi_{2\text{b}_i}(x)$ along with their energies, $E_{\text{bound}}^{i-}$. 
One can then in principle
obtain the Casimir energy using either one of the two equalities in Eq.\,(\ref{e2}), though the calculations could be very difficult and one might need to calculate the integrals numerically. 
In a previous paper \cite{dr} we considered a model described by the Lagrangian 
shown in Eq.\,(\ref{e1}) in which the pseudoscalar field $\phi(x)$ was prescribed and had a piecewise linear form.
That form of $\phi(x)$ along with the sine-Gordon soliton and the kink of the 
$\lambda \Phi^4$ model are shown in Fig.\,\ref{sgtoy}.
In this figure we indicate the parameters $\theta_0$, $\mu$ and $l$, which denote the value of $\phi(x)$ at $x=\infty$,
the slope of the pseudoscalar field at $x=0$ and the common scale of variation for the background fields, respectively.
The piecewise linear form chosen for $\phi(x)$ renders the problem exactly solvable and we are able to obtain the whole spectrum of the fermion \cite{dr}.
We refer to this system as the Simple Exactly Solvable Model (SESM). 
For that problem we have calculated the Casimir energy by the use of the direct approach \cite{la}.
However, for the problem with the sine-Gordon soliton, the equations of motion are not analytically solvable.
We obtain the spectrum of this system by the use of appropriate numerical methods. Throughout the paper we compare the results of the three models, wherever possible.
\par
It is worth noticing that the expressions given for the Casimir energy in Eq.\,(\ref{e2}) only contain the negative energy states.
However, in \cite{la} we showed that the total negative and positive energy densities are exactly the mirror images of each other for the SESM.
Therefore, the Casimir energy for that model can be computed only from the negative states or only the positive states,
or the average of all of the states and the results are exactly the same in all cases.
Notice that since all the symmetries of the model with the sine-Gordon soliton are the same as SESM,
the aforementioned argument is also true for the present model.
\section{The spectrum of the fermion}
The spectrum of a fermion is usually distorted due to its coupling to a background field.
In the presence of the background field spectral deficiencies can develope in the continua, and also the bound states can appear.
To obtain the distorted fermionic states and their energies in the presence of the prescribed sine-Gordon soliton,
we use the Dirac equation obtained from the Lagrangian (\ref{e1}).
Choosing the representation $\gamma^0=\sigma_1$, $\gamma^1=i\sigma_3$ and $\gamma^5=\gamma^0\gamma^1=\sigma_2$ for the Dirac matrices,
this equation becomes
\begin{equation}\label{e3}
i\sigma_1\partial_t\psi - \sigma_3\partial_x\psi - M \left[\cos
\phi\left(x,t\right)+i\sigma_2 \sin \phi\left(x,t\right)\right]\psi
=0,
\end{equation}
where $\psi=\left(\! \begin{array}{c}
\psi_{1}\\
\psi_2
\end{array}\! \right)$.
Using the definition $
\xi(x,t)=\left(\! \begin{array}{c}\psi_1+i\psi_2\\
\psi_1-i\psi_2\end{array}\! \right)=e^{-iEt} \left(\! \begin{array}{c}\xi_1(x)\\
\xi_2(x)\end{array}\! \right)$, the equations obeyed by $\xi_1(x)$ and $\xi_2(x)$ can be written in the following matrix form
\begin{equation}\label{eq}
 \left(\! \begin{array}{cc}i\partial_x-E&\, i M \mathrm{e}^{i \phi(x)}\\
-i M \mathrm{e}^{-i \phi(x)}&\, -i\partial_x-E\end{array}\! \right) \left(\! \begin{array}{c}\xi_1\\
\xi_2\end{array}\! \right)= \left(\! \begin{array}{c}0\\
0\end{array}\! \right).
\end{equation}
To obtain the fermionic bound states and their energies, we use a numerical method called the relaxation method.
This method is used for solving boundary value problems and it determines the solution by starting with a guess and improving it, iteratively.
We have already used this method for a system consisting of a fermion and a pseudoscalar field, where both fields were considered to be dynamical (not prescribed) \cite{leila}. The fermionic sector and the interaction part were identical to the Lagrangian given in 
Eq.\,(\ref{e1}) and the pseudoscalar field had dynamics given by
the $\lambda \Phi^4$ model with boundary conditions appropriate for a topologically nontrivial solution.
That is, we obtained simultaneously the exact fermionic bound states, their energies and the profile of the solitary wave, where the zero-order approximation of the latter is the kink. 
In the present problem the number of the coupled first-order ODEs is fewer by two, since the background field has no dynamics.
In our problem we use the fermionic bound states and their energies for the SESM \cite{dr} as initial guesses.
When using numerical methods, we rescale all the parameters with respect to the mass of the fermion ($M$), for simplicity.
\par
\begin{center}
\begin{figure}[th] \hspace{1.cm}\includegraphics[width=13.cm]{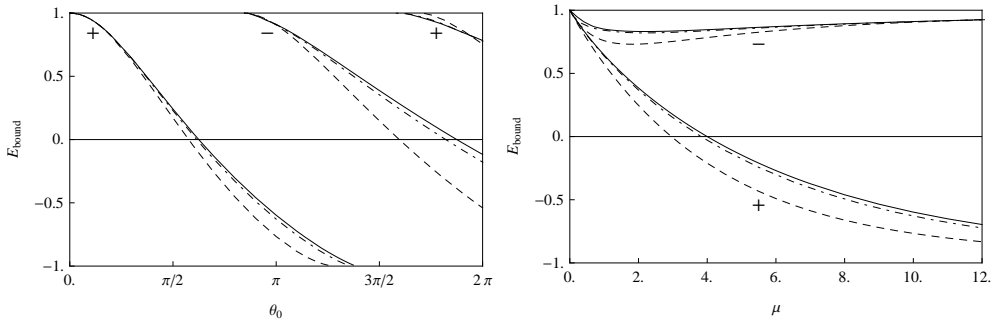}\caption{\label{bnd} \small
The energies of the bound states of the fermion. The left graph shows the bound state energies as a function of $\theta_0$ at $\mu=10$ and the right graph shows the bound state energies as a function of $\mu$ at $\theta_0=\pi$.
In both graphs solid, dashed and dotdashed lines are for the models with sine-Gordon soliton, SESM and the kink, respectively.
The $\pm$ signs indicate the parity of each bound state.}
\end{figure}
\end{center}

Figure \ref{bnd} shows the bound state energies obtained from the numerical results.
The left graph shows the bound energies as a function of $\theta_0$ when $\mu=10$. The right graph shows the bound state energies of the fermion as a function of $\mu$
at $\theta_0=\pi$, i.e.\;a soliton with winding number one.
In both graphs we also depict the bound state energies of the fermion for SESM and the model with the kink, for comparison.
\par
Now, we focus on the continuum states.
We first obtain the fermion continuum wave functions for a scattering process in which a plane wave enters the scattering region from the left or right.
Then, using the scattering wave functions, we construct the parity eigenfunctions which satisfy the parity condition $P\xi^{\pm}(x, t) =-\sigma_2\xi^{\pm}(−x, t)=\pm \xi^{\pm}(-x,t)$.
\par
The equations of motion for the real and imaginary parts of $\xi_1(x)=\eta_1(x)+i \eta_2(x)$ and $\xi_2(x)=\eta_3(x)+i \eta_4(x)$,
can be obtained from Eq.\,(\ref{eq}) and are as follows
\begin{align}\label{eqs}
&\eta^{'}_1+\cos \phi\left(x\right)\eta_3-E \eta_2-\sin \phi\left(x\right)\eta_4=0,\\\label{e71}&\eta^{'}_2+\cos \phi\left(x\right)\eta_4+E \eta_1+\sin
\phi\left(x\right)\eta_3=0,\\\label{e72}
&\eta^{'}_3+\cos \phi\left(x\right)\eta_1+E \eta_4+\sin
\phi\left(x\right)\eta_2=0,\\\label{e74} &\eta^{'}_4+\cos
\phi\left(x\right)\eta_2-E \eta_3-\sin \phi\left(x\right)\eta_1=0.
\end{align}
We cannot solve this set of equations analytically for the form chosen for $\phi(x)$, 
i.e.\;the sine-Gordon soliton.
Therefore, we use an appropriate numerical method to find the scattering wave functions.
We solve this set as an initial value problem and find the wave functions for the whole interval $(-\infty,+\infty)$,
using the Runge-Kutta-Fehlberg method of order six.
Since the form chosen for $\phi(x)$ in SESM is similar to the sine-Gordon soliton and is identical to it for $|x|\gg l$,
where $l$ is the scale of variation of these fields shown in Fig.\,\ref{sgtoy},
we can take advantage of the solutions of SESM to determine the initial boundary values for solving the equations.
We already have all the solutions of this model, including the scattering wave functions \cite{dr,levinsondr}, and this allows us to immediately calculate $\eta_i$s for any set of parameters \{$\theta_{0}$, $\mu$, $k$\}, where $E=\pm \sqrt {k^2+M^2}$.
The initial conditions can be alternatively obtained by solving Eqs.\,(\ref{eqs}-\ref{e74}) for $x\gg l$ where $\phi(x)=\theta_0$.
\begin{center}
\begin{figure}[th] \hspace{2.cm}\includegraphics[width=11.cm]{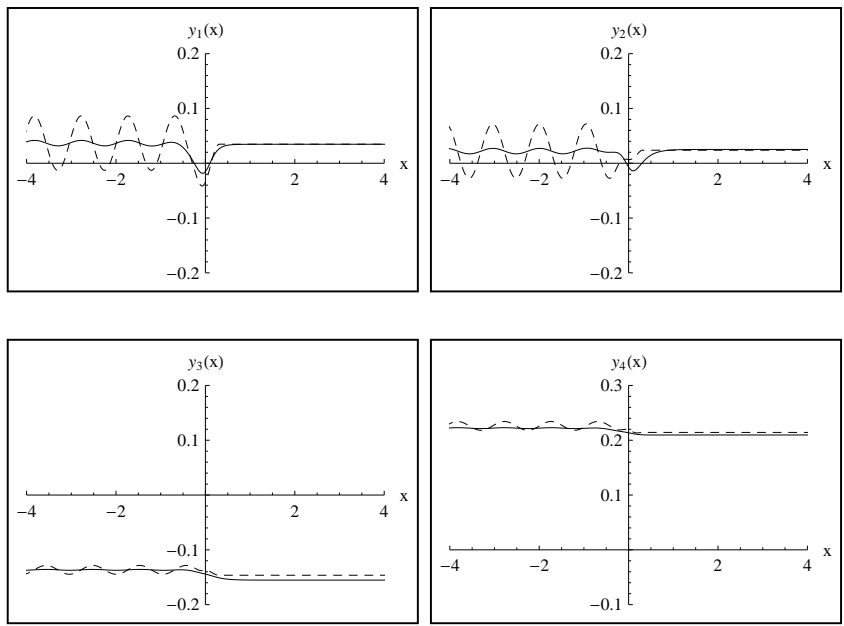}\caption{\label{scattering2} \small
The graphs of $y_i(x)$s ($=\eta_i(x)\textrm{e}^{-ikx}$) as functions of the spatial variable $x$, for the parameters $\theta_0=\pi$, $\mu=10$, $k=3$ and $E=+\sqrt{k^2+M^2}$.
Solid and dashed lines show $y_i(x)$s for our model and SESM, respectively. Note that the sine-Gordon model is almost reflectionless for the fermions.}
\end{figure}
\end{center}
\par
Since the set of equations (\ref{eq}) is a linear set for the Fermi field, any linear combination of its solutions is also a solution.
We take advantage of this fact to find the continuum parity eigenfunctions.
We denote the independent scattering solutions where the ``incident" wave is on the left (right) by $\xi_k^{\textrm{scat.}}(x,t)$
($\xi_{-k}^{\textrm{scat.}}(x,t)$).
We combine linearly these two solutions and determine the coefficients such that the combined solutions are simultaneous eigenstates of the Hamiltonian
and the parity operators. 
The results are
\begin{equation}
\xi_{k}^{\pm}(x,t)=\xi_{k}^{\textrm{scat.}}(x,t)\pm \xi_{-k}^{\textrm{scat.}}(x,t)= \mathrm{e}^{-iEt}
\left(\!\begin{array}{cc} \eta_1^{\pm}(x)\pm i\eta_2^{\pm}(x)\\ \eta_3^{\pm}(x)\pm i\eta_4^{\pm}(x)\end{array}\!\right)
=\mathrm{e}^{-iEt}\xi_{k}^{\pm}(x),
\end{equation}
 where the $\pm$ superscript denotes the parity. 
From the structure of this equation we can conclude that $\xi_{-k}^{\textrm{scat.}}(x,t)=P\xi_{k}^{\textrm{scat.}}(x,t)$, and a simple check verifies that this is indeed the case. 
For an alternative method of constructing the parity eigenstates see \cite{heavy3,heavy4}.
We can draw some interesting conclusions from these findings, once we factor out $\textrm{e}^{ikx}$ which is the common factor for the incoming and outgoing waves. 
Then, the oscillatory factor for reflected wave would be $\textrm{e}^{-2ikx}$. 
The results are shown in Fig.\,\ref{scattering2} whence we can immediately conclude that the sine-Gordon soliton, being completely reflectionless for elementary bosons, is almost reflectionless for the fermions. 
This property is also true for the kink \cite{la2}.
\section{The calculation of the Casimir energy using the phase shift method}
Now, having the whole spectrum of the system, we can calculate the Casimir energy and its parity decomposition and explore them in detail.
We shall use the phase shift method derived from the second part of Eq.\,(\ref{e2}) to calculate these energies.
\subsection{The phase shift method}
Our starting point is the following relation which holds for each of the positive and negative continua separately
\begin{equation}\label{ps}
\rho(k)-\rho_0(k)=\frac{1}{\pi}\frac{\mathrm{d}}{\mathrm{d}k}\delta(k).
\end{equation}
The quantity $\rho(k)-\rho_0(k)$ is the difference between the density of the continuum states in the free and interacting cases
and $\delta(k)$ denotes the scattering phase shift in the corresponding continuum.
Moreover, if the system possesses the parity symmetry, Eq.\,(\ref{ps}) holds for each parity channel in each of the continua, separately.
Substituting Eq.\,(\ref{ps}) into the second term of the second expression for the Casimir energy given in Eq.\,(\ref{e2}) and integrating by parts,
we obtain
\begin{align}\label{eps}
-&\int_{0}^{+\infty}\text{d}k\sqrt{k^2+M^2}\left[\rho^{\mathrm{sea}}(k)-\rho_0^{\mathrm{sea}}(k)\right]=
-\int_{0}^{+\infty}\frac{\text{d}k}{\pi}\sqrt{k^2+M^2}\frac{\mathrm{d}}{\mathrm{d}k}\left[\delta^{\mathrm{sea}}(k)-\delta^{\mathrm{sea}}(\infty)\right]\nonumber\\=&\int_{0}^{+\infty}\frac{\text{d}k}{\pi}\frac{k}{\sqrt{k^2+M^2}}\left[\delta^{\mathrm{sea}}(k)-\delta^{\mathrm{sea}}(\infty)\right]+\frac{1}{\pi}M\left[\delta^{\mathrm{sea}}(0)-\delta^{\mathrm{sea}}(\infty)\right].
\end{align}
In the first equality we have subtracted a zero term ($\frac{\mathrm{d}}{\mathrm{d}k}\delta^{\mathrm{sea}}(+\infty)$) from the original one.
Now we can compute the second term in the expression of the Casimir energy using the phase shifts.
\par
\begin{center}
\begin{figure}[th] \hspace{1.cm}\includegraphics[width=13.cm]{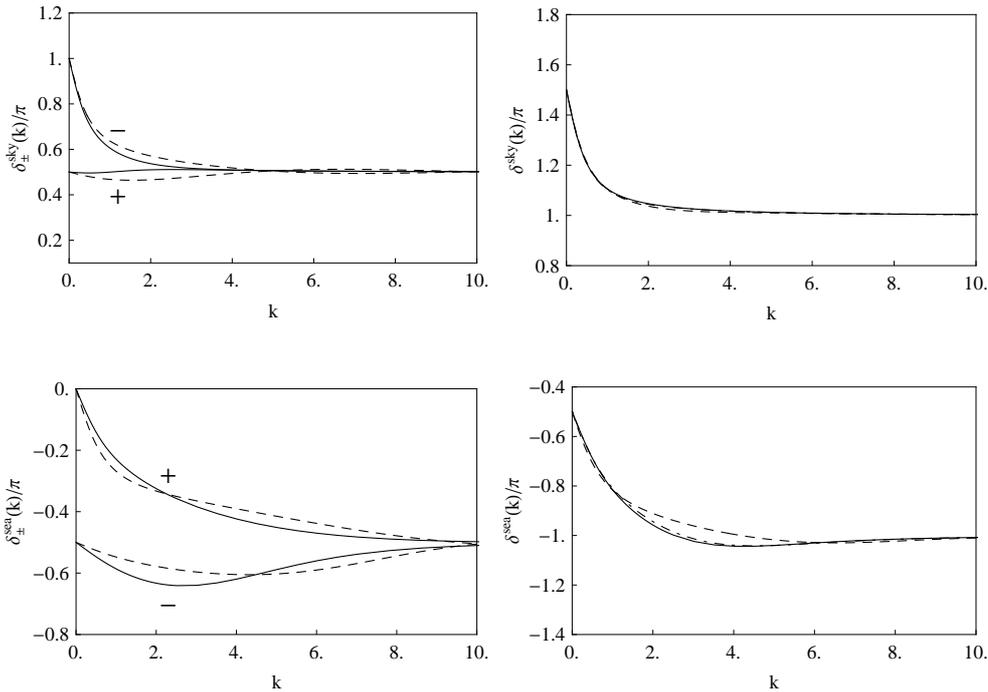}\caption{\label{figps} \small
The graphical representation of phase shifts of the continuum states with positive and negative energies, for $\mu=10$ and $\theta_0=\pi$.
The left graphs: The phase shift of the $\pm$ parity eigenstates, $\delta^{\mathrm{sky}}_{\pm}(k)$ and $\delta^{\mathrm{sea}}_{\pm}(k)$. The right graphs: The phase shifts of the scattering states,
$\delta^{\mathrm{sky}}(k)$ and $\delta^{\mathrm{sea}}(k)$. 
The $\pm$ signs in the left graphs indicate the parity of the corresponding eigenstates.
In these graphs the results for the models with the sine-Gordon soliton, SESM and the kink are shown by solid, dashed and dotdashed lines,
respectively. 
The phase shifts for the kink are available only for the scattering states, 
and as is apparent from the right graphs, they are very close to those of the sine-Gordon model.}
\end{figure}
\end{center}

Comparing the coefficients of $\mathrm{e}^{i k x}$ on the left- and right-hand sides of the scattering region, we can obtain the scattering matrix element, which is related to the scattering phase shift as $S(k)=\mathrm{e}^{i \delta(k)}$.
An analogous relation holds for each parity channel, separately, i.e.\;$S_{\pm}(k)=\mathrm{e}^{2 i \delta_{\pm}(k)}$. 
However, it is well known that the phase shifts of the upper and lower components are generically not equal, and this is indeed the case when the Fermi field is coupled to a topologically nontrivial configuration. 
We have devised a simple prescription to define a unique phase shift for the Dirac spinor \cite{levinsondr} and have tested its validity using both the strong and weak forms of the Levinson theorem \cite{la2,levinsondr,farid}. 
We shall use this prescription in this paper, and this will be yet another test. 
In Fig.\,\ref{figps} we plot the phase shifts for our system as a function of $k$, for the parameters $\theta_0=\pi$ and $\mu=10$. 
The right graphs show the $\delta^{\mathrm{sky}}(k)$ and $\delta^{\mathrm{sea}}(k)$, i.e.\;the phase shift for the scattering states with the positive and negative energies $\pm \sqrt{k^2+M^2}$ and the left graphs show $\delta^{\mathrm{sky}}_{\pm}(k)$ and $\delta^{\mathrm{sea}}_{\pm}(k)$, i.e.\;the phase shift for both parity eigenstates with the positive and negative energies. 
In these graphs we also depict the phase shift of SESM and the kink with the same parameters, for comparison.
\par
It is easy to check that the phase shifts depicted in Fig.\,\ref{figps} are consistent with both the weak \cite{levinson} and strong \cite{levinsondr} forms of the Levinson theorem.
Moreover, $\delta^{\mathrm{sea}}(k\rightarrow+ \infty)=- \theta_0$ and
$\delta^{\mathrm{sea}}_{\pm}(k\rightarrow+ \infty)=- \theta_0/2$,
which are consistent with the results of the adiabatic method of Goldstone and Wilczek \cite{goldstone}.

\subsection{The Casimir energy}
\begin{center}
\begin{figure}[th] \hspace{1.cm}\includegraphics[width=13.cm]{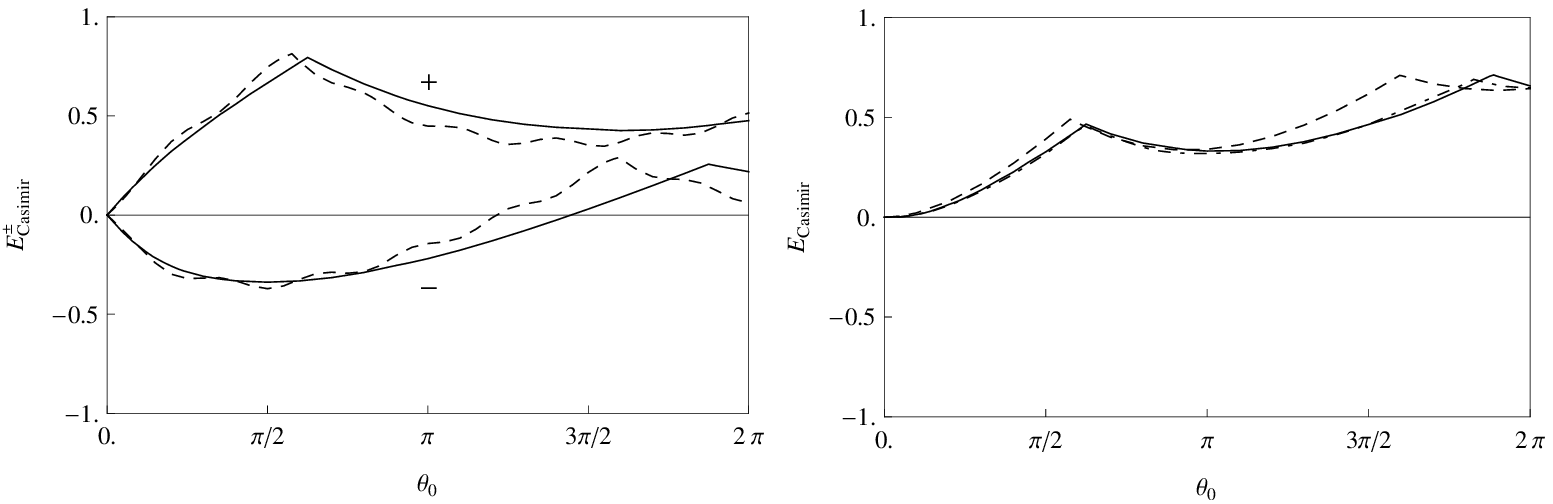}\caption{\label{casmu10} \small
 The graphical representation of the Casimir energy as a function of $\theta_0$, the value of the soliton at spatial infinity, at $\mu=10$.
 The left graph shows the Casimir energy for the $\pm$ parity channels separately for the models with the sine-Gordon soliton and SESM with solid and dashed lines, respectively.
The right graph shows the total Casimir energy for the models with the sine-Gordon soliton, SESM and the kink with solid, dashed and dotdashed lines, respectively.
}
\end{figure}
\end{center}

We now use Eqs.\,(\ref{e2}) and (\ref{eps}) and the information about the bound states shown in Fig.\,\ref{bnd} to calculate the Casimir energy and its parity decomposition for
different values of the parameters of the prescribed sine-Gordon soliton.
In Fig.\,\ref{casmu10} we present the Casimir energy as a function of $\theta_0$ for $\mu=10$ for the model with the sine-Gordon soliton, SESM and the kink,
by the solid, dashed and dotdashed lines, respectively.
As before, the left and right graphs in this figure show the Casimir energy for each parity channel and the total Casimir energy, respectively.
As can be seen, the total Casimir energy is, on the average, an increasing function of $\theta_0$ for all three models and there are
two mild maxima in the form of cusps in each graph.
Comparing these graphs with the left graph of Fig.\,\ref{bnd}, we conclude that these maxima occur when the bound state energies cross the line of $E=0$.
For our model in the interval $0\leqslant\theta_0\leqslant 2\pi$ a positive parity bound state crosses $E=0$ at $\theta_0\approx 0.625\pi$
and a negative parity one crosses $E=0$ at $\theta_0\approx 1.875\pi$.
The corresponding values for the SESM are $\theta_0\approx 0.576\pi$ and $\theta_0\approx 1.596\pi$.
From the left graph of Fig.\,\ref{casmu10} it can be seen that the Casimir energy for both parities has a mild cusp occurring where the levels
with the corresponding parity cross the line of $E=0$, as expected.
As before, the total Casimir energy in the right graph is the sum of the Casimir energies for $\pm$ parities in the left graph.
Notice that the Casimir energy in each parity channel for the SESM contains small amplitude oscillations, while those of the sine-Gordon soliton are smooth.
These oscillations are repercussions of the sharp edges of the pseudoscalar field of the SESM model shown in Fig.\,\ref{sgtoy}.
We have checked this conjecture by plotting these graphs for the same value of $\theta_0$ ($=\pi$) but for a smaller value of $\mu$, thus reducing the sharpness of the edges.
In that case the amplitude of oscillations becomes smaller and their ``wavelength'' becomes longer.
It is interesting to note that in either case the total Casimir energy is devoid of such oscillations.
\par
\begin{center}
\begin{figure}[th] \hspace{1.cm}\includegraphics[width=13.cm]{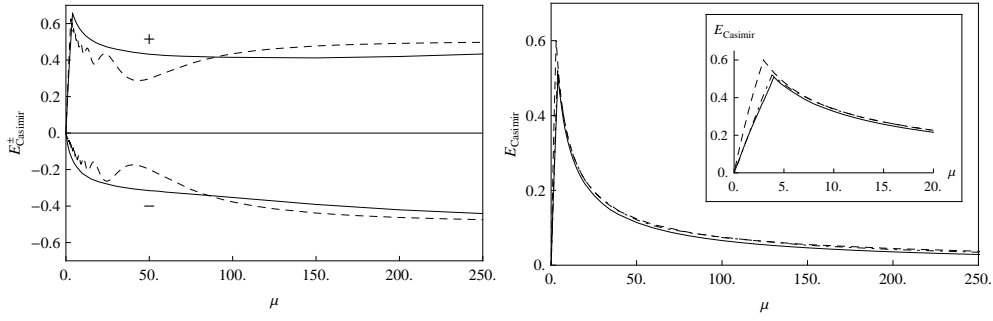}\caption{\label{casthpi} \small
The graphical representation of the Casimir energy as a function of $\mu$, the slope of the soliton profile at $x=0$, for $\theta_0=\pi$.
 The left graph shows the Casimir energy for the $\pm$ parity channels separately for the models with the sine-Gordon soliton and SESM with solid and dashed lines, respectively.
The right graph shows the total Casimir energy for the models with the sine-Gordon soliton, SESM and the kink with solid, dashed and dotdashed lines, respectively.
In the zoomed box we focus on small values of $\mu$ to show the details of the maximum and the differences between the results of all the three models.
}
\end{figure}
\end{center}

In Fig.\,\ref{casthpi} we plot the Casimir energy as a function of $\mu$ at $\theta_0=\pi$, i.e.\;a soliton with winding number 1,
for our model, SESM and the kink by the solid, dashed and dotdashed lines, respectively.
The left and right graphs of this figure show the Casimir energy for each parity and the total Casimir energy, respectively.
The parity of states in the left graph is denoted by the $\pm$ signs.
As can be seen from the right graph, in all three models, there is a sharp maximum occurring when the bound state energy crosses the line of $E=0$.
From the right part of Fig.\,\ref{bnd} we can see that for our model the positive parity bound state energy crosses the line of $E=0$ at $\mu \approx 4.000$,
while for SESM this crossing occurs at a lower value of the slope, i.e.\;$\mu\approx 2.957$.
The left graph of Fig.\,\ref{casthpi} confirms the fact that these bound states have positive parity.
Also, the value of the total Casimir energy is lower in the case of the sine-Gordon soliton as compared to SESM.
The largest difference between the graphs of these two models in the total Casimir energy occurs around the maximum, as is shown in the zoomed box of the right graph.
In all these models when the slope of $\phi(x)$ at $x=0$ decreases to zero, all of the Casimir energies approach zero, despite the residual nontrivial
boundary conditions.
Also, the total Casimir energy for all three models has the same limit when the slope of the pseudoscalar field tends to infinity.
This limit is zero at $\theta_0=\pi$, i.e.\;when we have a proper soliton with winding number one.
However, for other values of $\theta_0$ the Casimir energy is in general nonzero, when $\mu\rightarrow \infty$ in all three models (see Eq.\,(3.9) in \cite{la}). One can indeed calculate the total Casimir energy by using the scattering phase shift and the whole set of bound states of the fermion,
or equivalently by adding the Casimir energy obtained for each parity using the phase shift of the parity eigenstates and the corresponding bound states.
\par
We now explain the behavior of the Casimir energy in each parity channel, separately.
At $\theta_0=0$, i.e.\;the free case, there is a threshold bound state with positive parity at $E=+M$ and
one with negative parity at $E=-M$.
Therefore, at this point the spectral density in the Dirac sky (sea) is formally deficient by $1/2$ unit with positive (negative) parity.
As $\theta_0$ increases infinitesimally, one full positive parity bound state separates from the Dirac sky,
and the negative parity half-bound state sinks into the Dirac sea (see Fig.\,\ref{bnd}).
As a result, for an infinitesimal value of $\theta_0$, the Dirac sky has one unit of spectral deficiency with positive parity and the Dirac sea has no spectral deficiency.
As $\theta_0$ increases, Goldstone and Wilczek \cite{goldstone} showed that one unit of spectral deficiency develops in the Dirac sea when $\theta_0\rightarrow \pi$ due to the ever-present adiabatic contribution to the vacuum polarization.
This one unit of spectral deficiency is comprised of $1/2$ positive parity and $1/2$ negative parity contributions.
Analogously, one unit of adiabatic spectral surplus develops in the Dirac sky \cite{dr} as $\theta_0\rightarrow \pi$,
alongside the $1/2$ unit of positive parity spectral deficiency already present.
Now let us concentrate on the positive parity contribution to the Casimir energy shown in the left part of Fig.\,\ref{casmu10}. As $\theta_0\rightarrow \pi$ half a unit of positive parity spectral deficiency develops in the Dirac sea and the corresponding Casimir energy increases from zero.
At $\theta_0=0.625\pi$ the positive parity bound state crosses $E=0$ and starts reducing the Casimir energy.
This explains the cusp in Fig.\,\ref{casmu10}.
As $\theta_0$ increases further towards $\pi$, these two contributions continue to produce counteracting effects, and the negative slope after the cusp indicates that the effect of the full bound state descending dominates the developing $1/2$ unit of spectral deficiency.
Next, we analyze the behavior of the negative parity contribution to the Casimir energy.
As shown in reference \cite{dehghan}, when a bound state enters a continuum, the spectral distortion that is produces on the average ascends in the momentum space.
This explains the initial negative slope of the negative parity Casimir energy.
On the other hand the negative parity adiabatic spectral deficiency developing in the Dirac sea produces a positive contribution to the Casimir energy.
The graph of the Casimir energy shows that this effect dominates the former for $\theta_0\gtrsim \pi/2$.
Close to $\theta_0=2\pi$ a second bound state with negative parity crosses the line $E=0$, producing a cusp in the graph.
The total Casimir energy shown on the right part of Fig.\,\ref{casmu10} is the sum of two parity contributions.
Analogous explanations can be made for the results shown in Fig.\,\ref{casthpi}, using the fact that the positive parity bound state which joins the Dirac sea for $\theta_0\gtrsim \pi$ and the negative parity bound state which separates from the Dirac sky for $\theta_0\lesssim \pi$ when $\mu=10$, both become threshold bound states at $\theta_0=\pi$, as $\mu\rightarrow \infty$. 
An interesting conclusion which can be observed in Fig.\,\ref{casthpi} is that although the total Casimir energy goes to zero as $\mu\rightarrow \infty$, the individual parity contributions do not.
\section{Stability of the solutions}
\begin{center}
\begin{figure}[th] \hspace{1.cm}\includegraphics[width=13.cm]{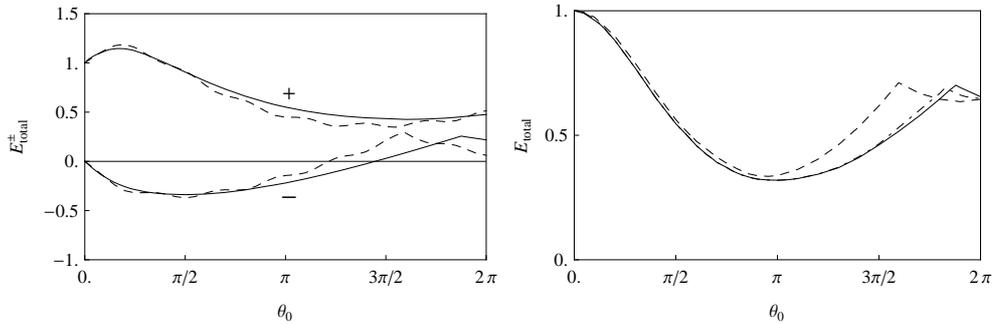}\caption{\label{stabilitymu10} \small
The left graph shows the total energy (the sum of the energy of a valence fermion in the lowest bound state and the Casimir energy) for positive and negative parities, separately, as a function of $\theta_0$ when $\mu=10$.
 The right graph shows the total energy.
Notice that the system attains its lowest energy at $\theta_0=\pi$.
Solid, dashed and dotdashed lines represent the results for the model with the sine-Gordon soliton, SESM and the kink, respectively.}
\end{figure}
\end{center}

\begin{center}
\begin{figure}[th] \hspace{1.cm}\includegraphics[width=13.cm]{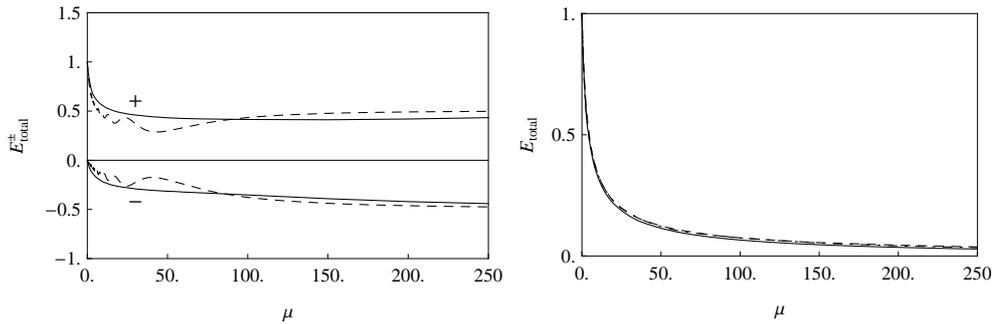}\caption{\label{stabilitythpi} \small
The left graph shows the total energy (the sum of the energy of a valence fermion in the lowest bound state and the Casimir energy) for positive and negative parities, separately, as a function of $\mu$ when $\theta_0=\pi$.
The right graph shows the total energy.
Solid, dashed and dotdashed lines represent the results for the model with the sine-Gordon soliton,  SESM and the kink, respectively.}
\end{figure}
\end{center}

In this section we consider a system consisting of a valence fermion in the ground state and explore the effect of the Casimir energy on its total energy.
The total energy for such a system is the sum of the Casimir energy and the energy of the valence fermion.
Notice that the energy of the valence fermion should not be added when this energy is negative,
since it has already been taken into account in the Casimir energy.
In Fig.\,\ref{stabilitymu10} we plot the total energies along with their parity decomposition as a function of $\theta_0$ for fixed $\mu=10$.
Also, Fig.\,\ref{stabilitythpi} shows the total energies along with their parity decomposition as a function of the slope $\mu$ of the pseudoscalar fields while their asymptotic
values are fixed at $\theta_0=\pi$.
In both figures solid, dashed and dotdashed lines refer to the results for the model with the sine-Gordon soliton, SESM and the kink, respectively.
Using the total Casimir energy shown in the right graphs of Figs.\,\ref{stabilitymu10} and \ref{stabilitythpi},
we are able to explore the stability of this system. 
It should be mentioned that since the ground state of the fermion has positive parity,
the total energy for the negative parity has only one contribution coming from the Casimir energy for the negative parity eigenstates.
As can be seen from Fig.\,\ref{stabilitymu10}, all three models have minima occurring at $\theta_0\approx \pi$, which corresponds to a soliton
with winding number one.
This means that not only this configuration is energetically favorable, but also it is stable against small fluctuations in the parameters of
the background field when this field is a soliton with a proper winding number, as expected.
Note that this stability is due to the combined effects of both parities.
The graphs of the total energies as a function of $\mu$ show no particular preference of the system for that parameter.

\section{Conclusion}
We have computed the Casimir energy for a coupled fermion-pseudoscalar field in which the pseudoscalar field is prescribed and chosen
to be the sine-Gordon soliton. This model is not exactly solvable.
Therefore, we use the relaxation method and the Runge-Kutta-Fehlberg method to obtain the bound states and
continuum scattering wave functions of the fermion, respectively.
Since the model is invariant under the parity and its equations of motion are linear, the continuum parity eigenfunctions of the system can be
obtained using a linear combination of the continuum scattering states and we have presented a very simple procedure to accomplish this.
The coefficients of this combination are determined by the use of the parity condition.
Having the complete spectrum of the system, including parity eigenstates, we have used the phase shift method to compute the total Casimir energy,
as well as the Casimir energy for each parity separately.
We have used a prescription we had introduced earlier for calculating a unique phase shift for the spinor, and to ascertain the resulting phase shifts, we have checked their consistency using the weak and strong forms of the Levinson theorem for each parity and the scattering phase shifts.
Moreover, we compute and plot the Casimir energy as a function of the slope of the pseudoscalar field at $x=0$ ($\mu$) and
its value at infinity ($\theta_0$). 
The graph of the Casimir energy as a function of $\mu$ has a sharp maximum at the value of $\mu$
in which the fermion has a zero-mode and goes to zero for $\mu \rightarrow 0$, and also for $\mu \rightarrow \infty$ when $\theta_0=n\pi$.
In the graph of the Casimir energy as a function of $\theta_0$, there are cusps whenever a fermion bound state energy level crosses the line of $E=0$.
Also, we calculate and plot the graph of the Casimir energy as a function of $\mu$ and $\theta_0$ for each parity to investigate what exactly happens
in the system by comparing them with the pattern of fermionic bound energy levels.
The Casimir energy is the sum of the Casimir energies for each parity.
We have also tested our method using a Simple Exactly Solvable Model (SESM).
Throughout the paper we compare our results with SESM and a similar model with the kink of the $\lambda \Phi^4$ theory, although in the latter case only results for total quantities are available and not their parity decomposition. 
The graphs of the Casimir energy for each parity for SESM are oscillatory while the ones of the model with sine-Gordon are smooth. This oscillatory behavior originates from the sharp edges of the background field of SESM.
In either case the total Casimir energy has no such oscillatory behavior. Moreover, we study the stability of the system in the presence of a valence fermion in the ground state. We show that the total energy has a minimum,
independent of the details of the shape of the background field, for a proper soliton with winding number one. Finally, we can state that, ignoring the aforementioned fine details, the qualitative behavior of the three models, i.e.\;the sine-Gordon model, the kink model and the SESM, are similar. The order with which we have stated the three models in the previous sentence, also represent the degree of distortion that these background fields cause in the spectrum of the Fermi field in ascending order. However, we should mention that the effects of the sine-Gordon model and the kink model are very close.

\section*{Acknowledgement} We would like to thank the research office
of the Shahid Beheshti University for financial support.


 \end{document}